\documentclass[aps,prd,superscriptaddress,amsmath,groupedaddress,       
         twocolumn,showpacs]{revtex4}
\usepackage{amssymb,graphicx}
\usepackage{dcolumn}

\newcommand{\msb}{{\rm\overline{MS}}}
\newcommand{\eq}[1]{Eq.~(\ref{#1})}
\newcommand{\order}{{\cal O}}
\newcommand{\xv}{{\bf x}}
\newcommand{\psib}{\overline{\psi}}
\newcommand{\g}{\gamma}
\newcommand{\lmc}{l_{ m_c}}
\newcommand{\amc}{am_{0c}}

\newcommand{\tdash}{\multicolumn{1}{c}{--}}
\newcommand{\mcmu}{0.986\,(10)}
\newcommand{\mcmc}{1.268\,(9)}
\newcommand{\mcmcpm}{1.268\pm0.009}
\newcommand{\almz}{0.1174\,(12)}
\newcommand{\almu}{0.251\,(6)}

\begin{document}

\title{High-Precision Charm-Quark Mass and QCD Coupling\\
        from Current-Current Correlators in Lattice and Continuum QCD}
\author{I.\ Allison}
\affiliation{TRIUMF, 4004 Wesbrook Mall, Vancouver, BC, V6T 2A3, 
Canada}
\author{E.\ Dalgic}
\affiliation{Physics Department, Simon Fraser University, Vancouver, 
British Columbia, Canada}
\author{C.\ T.\ H.\ Davies}
\affiliation{Department of Physics and Astronomy, University of 
Glasgow, Glasgow G12 8QQ, UK}
\author{E.\ Follana}
\affiliation{Physics Department, The Ohio State University, Columbus, 
Ohio 43210, USA}
\author{R.\ R.\ Horgan}
\affiliation{
Department of Applied Mathematics and Theoretical Physics, Cambridge 
University, Wilberforce Road, Cambridge CB3 0WA, UK}
\author{K.\ Hornbostel}
\affiliation{Southern Methodist University, Dallas, Texas 75275, USA}
\author{G.\ P.\ Lepage}
\email{g.p.lepage@cornell.edu}
\affiliation{Laboratory for Elementary-Particle Physics, Cornell 
University, Ithaca, NY 14853, USA}
\author{C.\ McNeile}
\affiliation{Department of Physics and Astronomy, University of 
Glasgow, Glasgow G12 8QQ, UK}
\author{J.\ Shigemitsu}
\affiliation{Physics Department, The Ohio State University, Columbus, 
Ohio 43210, USA}
\author{H.\ Trottier}
\affiliation{Physics Department, Simon Fraser University, Vancouver, 
British Columbia, Canada}
\author{R.\ M.\ Woloshyn}
\affiliation{TRIUMF, 4004 Wesbrook Mall, Vancouver, BC, V6T 2A3, 
Canada}
\collaboration{HPQCD Collaboration}
\noaffiliation

\author{K.\ G.\ Chetyrkin}
\author{J.\ H.\ K\"uhn}
\author{M.\ Steinhauser}
\affiliation{Institut f\"ur Theoretische Teilchenphysik,
Universit\"at Karlsruhe, D-76128 Karlsruhe, Germany}
\author{C.\ Sturm}
\affiliation{Physics Department,
Brookhaven National Laboratory,
Upton, New York 11973, U.S.A.
}
\date{April 24, 2008}
\preprint{	BNL-HET-08/11}
\preprint{	TTP08-18}
\preprint{	SFB/CPP-08-28}

\begin{abstract}
We use lattice QCD simulations, with MILC gluon configurations
and HISQ $c$-quark propagators,
to make very precise determinations of moments
of charm-quark pseudoscalar, vector and axial-vector correlators.
These moments are combined with new four-loop results from continuum 
perturbation
theory to obtain several new determinations of the $\msb$ mass of the 
charm quark and of
the $\msb$ coupling. We find
$m_c(3\,\mathrm{GeV})=\mcmu$\,GeV, or, equivalently,
$m_c(m_c)=\mcmc$\,GeV, both for $n_f=4$ flavors; and
$\alpha_\msb(3\,\mathrm{GeV},n_f\!=\!4)=\almu$, or, equivalently,
$\alpha_\msb(M_Z,n_f\!=\!5)=\almz$.
The new mass agrees well with results from
continuum analyses of the vector correlator using experimental data for
$e^+e^-$~annihilation (instead of using lattice QCD simulations). These 
lattice and
continuum results are the most accurate determinations to date of this 
mass. Ours is also one of the most accurate determinations of the QCD 
coupling by any method.
\end{abstract}

\maketitle

\section{Introduction}
Precise values for the QCD coupling $\alpha_\msb$ and the charm quark's 
mass~$m_c$ are important for high-precision tests of the Standard 
Model. Some of the most accurate mass determinations currently come 
from zero-momentum moments of current-current correlators built from 
the $c$~quark's electromagnetic current (see, for example, 
\cite{Kuhn:2001dm,Kuhn:2007vp}). Low moments are perturbative and have 
long been known through three-loop 
order~\cite{Chetyrkin:1995ii,Chetyrkin:1996cf,Chetyrkin:1997mb}. New 
techniques have recently extended these results to much higher 
moments~\cite{Boughezal:2006uu,Maier:2007yn} and, in some cases, to 
four-loop order~\cite{Chetyrkin:2006xg,Boughezal:2006px,Maier:private}. 
These moments can be estimated nonperturbatively, using dispersion 
relations, from experimental data for the electron-positron 
annihilation cross section, $\sigma(e^+ e^-\to\gamma^*\to X)$. The 
$c$~quark's mass is extracted by comparing the perturbative and 
experimental determinations.

In this paper we show how to compute such moments directly using 
accurately tuned, highly realistic numerical simulations of QCD in the 
lattice approximation~\cite{Bochkarev:1995ai}. As we will show, the 
correlator moments obtained nonperturbatively from such simulations can 
be used in place of data from $e^+ e^-$~annihilation to obtain  new, 
percent-accurate determinations of the $c$~quark's mass. With lattice 
QCD, it is also possible to replace the electromagnetic current in the 
correlator by the pseudoscalar operator $m_c\psib_c\gamma_5\psi_c$, 
thereby providing a completely new set of mass determinations and an 
important cross check on the entire methodology.

In fact the pseudoscalar correlators are particularly easy to simulate, 
and there are no renormalization factors required since the 
corresponding axial-vector current is partially conserved in our 
lattice formalism. Consequently these correlators give our most 
accurate masses.  They also give very accurate values for the QCD 
coupling, when combined with our new four-loop results from 
perturbation theory.

In Section~\ref{corr-from-latt} we describe how to compute pseudoscalar 
correlators and their moments using lattice QCD. We discuss techniques 
for reducing lattice artifacts in Section~\ref{sec:qcd_simulations}, 
and present new determinations of the $c$-quark mass and QCD coupling 
from our lattice ``data'' in Section~\ref{sec:mc}. In 
Section~\ref{sec:other} we extend our analysis to include vector and 
axial-vector correlators. We summarize our main results in 
Section~\ref{concl}. In the Appendix we review the continuum 
perturbation theory needed for this analysis, including new four-loop 
results for the pseudoscalar and vector cases.

\section{Lattice QCD and Pseudoscalar 
Correlators}\label{corr-from-latt}
Few-percent accurate QCD simulations have only become possible in the 
last few years (see, for example,~\cite{Davies:2003ik,Mason:2005zx}), 
and accurate simulations of relativistic $c$~quarks only in the past 
year\,---\,with the new Highly Improved Staggered Quark (HISQ) 
discretization of the quark 
action~\cite{Follana:2006rc,Follana:2007uv}, which we use here. A 
lattice QCD simulation proceeds in two steps. First the QCD 
parameters\,---\,the bare coupling constant and bare quark masses in 
the Lagrangian\,---\,must be tuned. Then the tuned simulation is used 
to compute vacuum matrix elements of various quantum operators from 
which physics is extracted. An obvious approach to the tuning is to 
choose a lattice spacing~$a$, and then tune each of the QCD parameters 
so that the simulation reproduces the experimental value for a 
corresponding physical quantity that is well measured. It is more 
efficient, however, to first choose a value for the bare coupling and 
then adjust the lattice spacing and bare masses to give physical 
results.

In the simulations used here, we set the lattice spacing to reproduce 
the correct $\Upsilon^\prime-\Upsilon$~meson mass difference in the 
simulations~\cite{Gray:2005ur}, while we tune the $u/d$, $s$, $c$ 
and~$b$ masses to give correct values for $m_\pi^2$, $2m^2_K-m_\pi^2$, 
$m_{\eta_c}$, and $m_\Upsilon$, respectively. (For efficiency we set 
$m_u=m_d$; this leads to negligible errors in the analysis presented 
here.) The important parameters for the particular simulations used in 
this paper are listed in Table~\ref{tab:param}; further details can be 
found in~\cite{Follana:2007uv,Davies:2003ik}. Once these parameters are 
set, there are no further physics parameters, and the simulation will 
accurately reproduce QCD physics for momenta much smaller than the 
ultraviolet~(UV) cutoff~($\Lambda\sim\pi/a$).

We have tested these simulations extensively (see, for example, 
\cite{Davies:2003ik,Mason:2005zx,Follana:2006rc,Follana:2007uv,Gray:2005ur}) and, in particular, we have done very precise tests for the charm-quark physics most relevant to this work. These demonstrate, for example, that our simulations reproduce the low-lying spectrum, including spin structure, of both charmonium and heavy-light mesons ($D$ and $D_s$) to within  our simulation uncertainties (a few percent or less)~\cite{Follana:2006rc,Follana:2007uv}.

\begin{table}
    \caption{Parameters for the QCD simulations used in this paper. The
    inverse lattice spacing $a^{-1}$ is in units of 
$r_1=0.321(5)$\,fm~\cite{Gray:2005ur},
	defined in terms of the static-quark potential~\cite{MILC}. $L$ and 
$T$ are the
	spatial and temporal size of the lattices used for each set of gluon 
configurations.
	The configurations used here were generated by the MILC
	collaboration~\cite{MILC} with $u$, $d$ and $s$ sea-quarks.
	The $u$~and $d$~masses are set equal to $m_{u/d}$. The sea-quark 
masses
	are given in the standard MILC notation which
	includes a factor of the (plaquette) tadpole factor $u_0$.
	}
    \label{tab:param}
\begin{center}
    \begin{ruledtabular}\begin{tabular}{cccccccc}
Set & $r_1/a$ & $au_0m_{0u/d}$ & $au_0m_{0s}$ & $am_{0c}$ & $u_0$ & 
$L/a$ & $T/a$ \\ \hline
1 & 2.133(14) & 0.0097 & 0.048 & 0.850 & 0.860 & 16 & 48  \\
2 & 2.129(12) & 0.0194 & 0.048 & 0.850 & 0.861 & 16 & 48  \\ \\
3 & 2.632(13) & 0.0050 & 0.050 & 0.650 & 0.868 & 24 & 64  \\
4 & 2.610(12) & 0.0100 & 0.050 & 0.660 & 0.868 & 20 & 64  \\
5 & 2.650(8) & 0.0200 & 0.050 & 0.648 & 0.869 & 20 & 64  \\ \\
6 & 3.684(12) & 0.0062 & 0.031 & 0.430 & 0.878 & 28 & 96  \\
7 & 3.711(13) & 0.0124 & 0.031 & 0.427 & 0.879 & 28 & 96  \\ \\
8 & 5.277(16) & 0.0036 & 0.018 & 0.280 & 0.888 & 48 & 144  \\
\end{tabular}\end{ruledtabular}
\end{center}
\end{table}

Given a tuned simulation, it is straightforward to calculate 
correlators of the sort used to determine~$m_c$. The simplest of these 
is for the $c$~quark's pseudoscalar density, 
$j_5\equiv{\psib}_{c}\gamma_5\psi_c$:
\begin{equation}
    G(t) \equiv a^6\,\sum_\xv (am_{0c})^2 \langle0| j_{5}(\xv,t) 
j_{5}(0,0)
    |0\rangle
\end{equation}
where $m_{0c}$ is the $c$~quark's bare mass (in the lattice 
Lagrangian). Here time~$t$ is euclidean, and the sum over spatial 
position~$\xv$ sets the total three momentum to zero. Note that 
$G(t)=G(T-t)=G(T+t)$ where $T$ is the temporal length of the lattice.

We include two factors of $am_{0c}$ in the definition of $G(t)$ so that 
$G(t)$ becomes independent of the UV cutoff as~$a\to 
0$~\cite{lepage:pcac}. Consequently the lattice and continuum $G(t)$s 
become equal in this limit. Moments $G_n$ are trivially computed:
\begin{equation} \label{Gn-def}
    G_n \equiv \sum_t (t/a)^n G(t),
\end{equation}
where, on our periodic lattice~\cite{footnote:0},
\begin{equation}
    t/a \in \{0,1,2\,\ldots\,T/2a-1,0,-T/2a+1\,\ldots\,-2,-1\}.
\end{equation}
The cutoff independence of $G(t)$ implies that
\begin{equation}\label{gn-def}
    G_n = \frac{g_n(\alpha_\msb(\mu),\mu/m_c)}{(am_c(\mu))^{n-4}} + 
\order((am_c)^m)
\end{equation}
for $n\ge4$, where $m_c(\mu)$ is the $\msb$ mass at scale~$\mu$ and 
$g_n$ is dimensionless. The $c$~mass can be determined from moments 
with $n\ge6$ given $G_n$ from lattice simulations and $g_n$ from 
perturbation theory (see Appendix), while the QCD coupling can be 
determined from the dimensionless moment~$G_4$. This assumes that 
perturbation theory is applicable, which should be the case for small 
enough~$n$.

Note that here and elsewhere in this paper we omit annihilation 
contributions from $c\bar{c}\to\mathrm{gluons}\to c\bar{c}$. This is 
allowed provided we omit the same contributions from perturbation 
theory. Annihilation contributions to the nonperturbative part of our 
analysis would be negligible (for example, they shift the $\eta_c$~mass 
by approximately~2.4\,MeV, which is less 
than~0.1\%~\cite{Follana:2006rc,lepage:annihilation}).

\section{QCD Simulations} \label{sec:qcd_simulations}
\subsection{Reduced Moments}\label{sec:red-mom}
The biggest challenge when using lattice QCD to produce $c$-quark 
correlator moments is controlling: 1) $\order((am_c)^n)$ errors caused 
by the lattice approximation; and 2)  tuning errors in the QCD 
parameters, and especially in the lattice spacing and the $c$-quark's 
bare mass. We reduce each of these sources of error by making two 
modifications to the moments.

First we replace $G_n$ by
\begin{equation}
	\frac{G_n}{G_n^{(0)}} = \frac{g_n}{g_n^{(0)}} \left(
	\frac{m_{\mathrm{pole},c}^{(0)}}{m_c(\mu)}\right)^{n-4} + 
\order((am_c)^m\alpha_s)
\end{equation}
where $G_n^{(0)}$ is the $n^\mathrm{th}$~moment of the correlator to 
lowest order in lattice QCD perturbation theory~\cite{lepage:tip}, and 
$g_n^{(0)}$ is the lowest-order part of $g_n$ in continuum perturbation 
theory. The lowest-order on-shell or ``pole'' mass of the $c$-quark 
sets the mass scale in the lowest-order lattice moments:
\begin{equation}
	G_n^{(0)} = \frac{g_n^{(0)}}{(am_{\mathrm{pole},c}^{(0)})^{(n-4)}} + 
\order((am_c)^m)
\end{equation}
In the HISQ formalism, this mass is related to the mass $m_{0c}$ that 
appears in the action by~\cite{Follana:2006rc}:
\begin{align}
	m_{\mathrm{pole},c}^{(0)} &=  m_{0c} \left(
	1-\frac{3\, (\amc)^4}{80} + \frac{23\,(\amc)^6}{2240} 
\right.\nonumber\\
	&+ \left.
	\frac{1783\,(\amc)^8}{537600} - \frac{76943\,(\amc)^{10}}{23654400} + 
\cdots\right).
\end{align}
Introducing $G_n^{(0)}$ removes the explicit factors of the lattice 
spacing in the denominator of~\eq{gn-def}, and also cancels finite-$a$ 
errors to all orders in~$a$ and zeroth order in~$\alpha_s$. Thus we 
expect finite-$a$ errors that are reduced by a factor of 
order~$\alpha_s(1/a)\approx1/3$ when we divide $G_n$ by the 
corresponding lowest-order lattice moment; and we find in practice that 
they are 3--4~times smaller.

A second modification is to replace the pole mass in $G_n/G_n^{(0)}$ by 
the value of the $\eta_c$ mass obtained from the simulation, 
$am_{\eta_c}$ (in lattice units)~\cite{footnote:1}:
\begin{equation}
	\frac{G_n}{G_n^{(0)}}\left(\frac{am_{\eta_c}}{2am_{\mathrm{pole},c}^{(0)}}
	\right)^{n-4} = \frac{g_n}{g_n^{(0)}} \left(
	\frac{m_{\eta_c}}{2m_c(\mu)}\right)^{n-4}
\end{equation}
up to $\order((am_c)^m\alpha_s)$ corrections. With this additional 
factor, the leading dependence on $m_c(\mu)$  enters through the ratio 
$m_c(\mu)/m_{\eta_c}$. Consequently small errors in the simulation 
parameter $am_{0c}$ are mostly cancelled in this expression by 
corresponding shifts in the simulation value for~$am_{\eta_c}$. This 
cancellation is accurate up to binding corrections of 
order~$(v_c/c)^2\approx1/3$ in~$m_{\eta_c}$, and therefore the impact 
of any tuning error in~$m_{0c}$ is three times smaller with this 
modification~\cite{footnote:2}.

Combining these two modifications, we replace $G_n$ by a reduced 
moment:
\begin{equation}\label{Rn-def}
    R_n \equiv \left\{
    \begin{aligned}
        & G_4/G^{(0)}_4 & & \text{for $n=4$,} \\
        &\frac{a m_{\eta_c}}{2a m_{\mathrm{pole},c}^{(0)}} 
\left(G_n/G^{(0)}_n\right)^{1/(n-4)}
        & & \text{for $n\ge6$,}
    \end{aligned}\right.
\end{equation}
The reduced moments can again be written in terms of continuum 
quantities:
\begin{equation}\label{Rn-eqn}
    R_n \equiv \left\{
	\begin{aligned}
		& r_4(\alpha_\msb,\mu/m_c) & & \text{for $n=4$,} \\
		& \frac{r_n(\alpha_\msb,\mu/m_c)}{2 m_c(\mu)/m_{\eta_c}}
		& & \text{for $n\ge6$,}
	\end{aligned}\right.
   \end{equation}
up to $\order((am_c)^m\alpha_s)$ corrections, where  $r_n$ is obtained 
from $g_n$ (\eq{gn-def}) and its value, $g_n^{(0)}$, in lowest-order 
continuum perturbation theory:
\begin{equation}\label{rn-def}
    r_n = \begin{cases} g_4/g_4^{(0)} & \text{for $n=4$,}\\
                \left(g_n/g_n^{(0)}\right)^{1/(n-4)} & \text{for 
$n\ge6$.}
        \end{cases}
\end{equation}

\begin{figure}
          \includegraphics[scale=1.0]{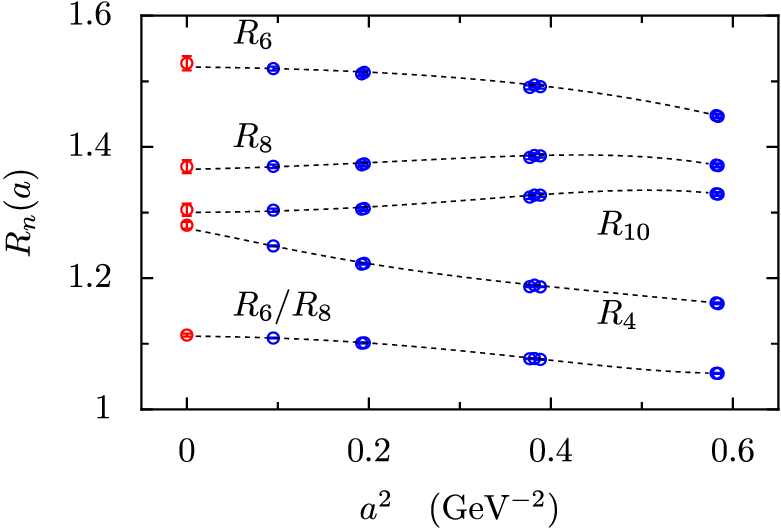}
       \caption{Reduced moments $R_n$ and a ratio of these moments from 
lattice simulations with different lattice spacings~$a$. The tight 
clusters of points at each of the three largest lattice spacings 
correspond to results for different sea-quark masses. The dashed lines 
show the functions used to fit the lattice results, with the sea-quark 
masses set equal to the masses used at the smallest lattice spacing. 
These extrapolation functions were used to obtain the~$a=0$, 
$m_{u/d/s}=0$~results shown in the plot.}
    \label{fig:rn}
\end{figure}

The $c$~mass is obtained from \eq{Rn-eqn} with $n\ge6$ using the 
nonperturbative lattice QCD (LQCD) value for $R_n$, the perturbative 
QCD (PQCD) estimate for $r_n$, and the experimental value for 
$m_{\eta_c}$,~2.980\,GeV:
\begin{equation}\label{extract-mc}
    m_c(\mu) = \frac{m_{\eta_c}^\mathrm{exp}}{2}\, 
\frac{r_n^\mathrm{PQCD}}{R_n^\mathrm{LQCD}}.
\end{equation}
Reduced moment $R_4$ is dimensionless and so depends only weakly on 
$m_c$. Simulation values for this moment can be compared with 
perturbation theory to obtain estimates for the QCD coupling: given the 
$c$-quark mass and a nonperturbative lattice QCD value for $R_4$, we 
solve the equation
\begin{equation}
	R_4^\mathrm{LQCD} = r_4(\alpha_\msb,\mu/m_c)
\end{equation}
for $\alpha_\msb(\mu)$.
Ratios of reduced moments, like $R_n/R_{n+2}$ for $n\ge6$, can also be 
used in this way to estimate the coupling.

\begin{table*}
    \caption{Simulation results for $R_n(a,m_{u/d},m_s)$ for different 
lattice
	parameter sets (see Table~\ref{tab:param}).
    The inverse lattice spacing~$a^{-1}$ is in~GeV. Extrapolations to 
zero lattice
	spacing and zero sea-quark masses are given for each quantity,
    together with the
    corresponding value for $m_c(\mu)$ (in GeV) or $\alpha_\msb(\mu)$
	for $n_f=4$ flavors and $\mu=3$\,GeV.}
    \label{tab:rn}
\begin{center}
    \begin{ruledtabular}\begin{tabular}{ccccccccc|cc}
Set: & 1 & 2 & 3 & 4 & 5 & 6 & 7 & 8  \\
$a^{-1}$: & 1.31 & 1.31 & 1.62 & 1.60 & 1.63 & 2.26 & 2.28 & 3.24 & 
$a,m_{u/d/s}\to0$ & $m_c(\mu)$ \\\hline
$R_{6}$ & 1.448(3) & 1.447(3) & 1.494(3) & 1.492(3) & 1.491(3) & 
1.514(3) & 1.511(3) & 1.519(3) & 1.528(11) & 0.986(10) \\
$R_{8}$ & 1.372(3) & 1.371(3) & 1.387(3) & 1.386(3) & 1.384(3) & 
1.374(3) & 1.373(3) & 1.370(3) & 1.370(10) & 0.986(11) \\
$R_{10}$ & 1.329(3) & 1.328(3) & 1.326(3) & 1.326(3) & 1.324(3) & 
1.306(3) & 1.305(3) & 1.304(3) & 1.304(9) & 0.973(19) \\
$R_{12}$ & 1.294(3) & 1.293(3) & 1.284(3) & 1.284(3) & 1.281(3) & 
1.263(3) & 1.262(3) & 1.262(3) & 1.265(9) & 0.969(23) \\
$R_{14}$ & 1.264(3) & 1.264(3) & 1.252(2) & 1.251(2) & 1.248(2) & 
1.232(2) & 1.231(2) & 1.232(2) & 1.237(9) & 0.967(28) \\
$R_{16}$ & 1.239(2) & 1.239(2) & 1.228(2) & 1.226(2) & 1.223(2) & 
1.207(2) & 1.206(2) & 1.210(2) & 1.215(9) & 0.965(33) \\
$R_{18}$ & 1.218(2) & 1.218(2) & 1.208(2) & 1.205(2) & 1.202(2) & 
1.187(2) & 1.187(2) & 1.191(2) & 1.198(9) & 0.963(38) \\
\end{tabular}\end{ruledtabular}\\[1ex]
	\begin{ruledtabular}\begin{tabular}{ccccccccc|cc}
Set: & 1 & 2 & 3 & 4 & 5 & 6 & 7 & 8  \\
$a^{-1}$: & 1.31 & 1.31 & 1.62 & 1.60 & 1.63 & 2.26 & 2.28 & 3.24 & 
$a,m_{u/d/s}\to0$ & $\alpha_\msb(\mu)$ \\\hline
$R_4$ & 1.162(1) & 1.161(1) & 1.189(1) & 1.187(1) & 1.187(1) & 1.223(1) 
& 1.221(1) & 1.249(1) & 1.281(5) & 0.252(6) \\
$R_{6}/R_{8}$ & 1.055(1) & 1.055(1) & 1.078(1) & 1.076(1) & 1.077(1) & 
1.101(1) & 1.101(1) & 1.109(1) & 1.113(2) & 0.249(6) \\
$R_{8}/R_{10}$ & 1.033(1) & 1.033(1) & 1.046(1) & 1.045(1) & 1.046(1) & 
1.052(1) & 1.052(1) & 1.051(1) & 1.049(2) & 0.224(31) \\
$R_{10}/R_{12}$ & 1.027(1) & 1.027(1) & 1.033(1) & 1.033(1) & 1.034(1) 
& 1.034(1) & 1.034(1) & 1.033(1) & 1.031(2) & 0.241(30) \\
$R_{12}/R_{14}$ & 1.023(1) & 1.023(1) & 1.025(1) & 1.026(1) & 1.026(1) 
& 1.025(1) & 1.025(1) & 1.024(1) & 1.022(2) & 0.243(47) \\
$R_{14}/R_{16}$ & 1.020(1) & 1.020(1) & 1.020(1) & 1.021(1) & 1.021(1) 
& 1.020(1) & 1.020(1) & 1.019(1) & 1.017(2) & 0.242(70) \\
$R_{16}/R_{18}$ & 1.017(1) & 1.017(1) & 1.016(1) & 1.017(1) & 1.017(1) 
& 1.017(1) & 1.017(1) & 1.016(1) & 1.014(2) & 0.241(96) \\
\end{tabular}\end{ruledtabular}
\end{center}
\end{table*}

\subsection{Simulation Results}
\label{sub:simulation_results}
Our simulation results for $R_n(a,m_{u/d},m_s)$ are listed for 
different moments~$n$, lattice spacings~$a$, and sea-quark masses in 
Table~\ref{tab:rn}; we also list ratios of reduced moments, 
$R_n/R_{n+2}$ for~$n\ge6$. Some of these results are plotted versus the 
lattice spacing in Figure~\ref{fig:rn}. Our simulations did not include 
$c$-quark vacuum polarization, but the correction to the moments can be 
computed using perturbation theory (since $c$-quarks are relatively 
heavy)~\cite{footnote:2.2}. We find that these corrections add 0.7\% to 
$R_4$, and are of order 0.1\% or less for the higher moments considered 
here. The $R_n$s in the table are corrected to include this effect.

The uncertainty quoted for each $R_n(a,m_{u/d},m_s)$ with $n\ge6$ is 
dominated by the uncertainty in our tuning of~$m_{0c}$ (other sources, 
such as statistical or finite volume errors~\cite{footnote:2.1}, are 
negligible). We tune the bare $c$-quark mass so that our simulations 
give correct masses for the $\eta_c$~\cite{footnote:1.1}. Our tuning is 
limited by the precision with which we know the lattice spacing~$a$ for 
any given parameter set in Table~\ref{tab:param}, since simulations 
give masses in lattice units (that is, $am_{\eta_c}$). We determine 
lattice spacings by combining MILC's values for $r_1/a$ (see 
Table~\ref{tab:param}), which are accurate to around~0.6\%, with a 
value for $r_1$ determined from the upsilon spectrum: 
$r_1=0.321(5)$\,fm~\cite{Gray:2005ur}, which is accurate to~1.5\%. The 
corresponding uncertainties in the $R_n$s are three times smaller, 
because of the reduced sensitivity of $m_c/m_{\eta_c}$ (see above). We 
include the uncertainty due to $r_1/a$ ($0.6\%/3=0.2\%$) in the 
uncertainties reported for each separate~$R_n(a)$ in 
Table~\ref{tab:rn}. The uncertainty due to $r_1$ ($1.5\%/3=0.5\%$)  is 
included only in our final results, after extrapolation (since changes 
in $r_1$ affect all $R_n(a)$s by the same amount).

Reduced moment~$R_4$ and our ratios of reduced moments are much less 
sensitive to  errors in~$m_{0c}$ because the $c$-mass enters only 
through radiative corrections, in perturbation theory. Consequently 
uncertainties due to mistunings of $m_{0c}$ are smaller by an order of 
magnitude or more for these quantities. We account for potential tuning 
errors by assigning an uncertainty of~0.05\% to each of the $R_4$s and 
ratios.

We also include in Table~\ref{tab:rn} our results extrapolated to zero 
lattice spacing and zero sea-quark mass~\cite{footnote:1.2}. Our 
extrapolation procedure is described in the next section.

\subsection{$am_c$, $m_q/m_c$ Extrapolations}
\label{amc_extrapolation}
Table~\ref{tab:rn} and Figure~\ref{fig:rn} show that our reduced 
moments depend only weakly on the lattice spacing, with most moments 
changing by 0.5\% or less between our two smallest lattice spacings and 
only a few percent over our entire range. The dependence on the 
sea-quark masses is even weaker. We nevertheless correct our results by 
fitting the variation over our different sets of lattice parameters 
(Table~\ref{tab:param})  and extrapolating to zero lattice spacing and 
zero sea-quark mass. We do this with a constrained 
fit~\cite{Lepage:2001ym,Davies:2008sw} of our simulation data to a 
function of the form:
\begin{align}\label{eq:extrap}
    R_n(a) = & R_n(0)\left(1   + c_{n,2}(am_c)^2\alpha_s
        + c_{n,4}(am_c)^4\alpha_s  \right.\nonumber \\
		&+ \left. c_{n,6} (am_c)^6\alpha_s + c_{n,8} (am_c)^8\alpha_s
		+ \cdots \right) \nonumber \\
		&\times \left( 1 + f_{n,1} (2m_{u/d}+m_s)/m_c + \cdots \right)
\end{align}
where we take~$m_c=1$\,GeV and $\alpha_s=\alpha_s(1/a)$. This form is 
motivated by the pattern of $a^n$ errors in our lattice actions, and by 
chiral perturbation theory, which implies nonperturbative corrections 
that depend linearly on the sea-quark masses. (There is sea-quark mass 
dependence in perturbation theory, as well, but it enters at 
$\order(\alpha_s^2 (m_q/m_c)^2)$ and is negligible here.)

The extrapolated results are largely independent of the exact 
functional form used for the extrapolation provided reasonable Bayesian 
priors are included (in the $\chi^2$ function that is minimized in the 
fit) for each of the coefficients~$c_{n,i}$ and 
$f_{n,i}$~\cite{Lepage:2001ym,Davies:2008sw}. We use the same Gaussian 
prior, centered at zero with width~$\sigma_c=1$, for every $c_{n,i}$, 
for all moments and moment ratios except~$R_4$. Moment $R_4$ has larger 
$a^2$ errors and needs a wider prior; we take $\sigma_c=5$. We use a 
prior with width $\sigma_f=0.1$ for the $f_{n,i}$, which is twice as 
large as the largest coefficient obtained from the fits.

The error estimates from our fits initially increase, and $\chi^2$ 
decreases, as we add successively higher-order terms in the $(am_c)^2$ 
and $m_q/m_c$ expansions. Eventually the errors stop increasing and 
$\chi^2$ stops changing, again assuming proper priors for all fit 
parameters. It is important to add terms through this point in order to 
avoid underestimating uncertainties in the final fit results. Adding 
further terms has no effect on fit results (values or errors).

Only a single term is needed in each series to get good fits for $R_n$ 
with $n\ge6$ if we discard data from the largest lattice spacing; and 
our final results (values and errors) are little changed. We, however, 
retain results from the coarsest lattices, despite the large value of 
$am_c$ for those lattices, in order to test our priors. Fitting all of 
our simulation data, we get good fits with two terms in the $(am_c)^2$ 
and a single term in the $m_q/m_c$ expansion. To be certain of 
convergence, we used eight terms in the first expansion and two in the 
second to obtain the extrapolated results in Table~\ref{tab:rn}. The 
fact that we get good fits ($\chi^2$ per data point less than one) even 
when we include data from the coarsest lattices helps validate the 
design of our fit function and priors, and it reassures us that our 
fits are not underestimating errors.

Moment $R_4$ and the ratios of moments are more accurately determined 
in our simulation than the other $R_n$s, and so typically require an 
additional term in the $(am_c)^2$ expansion. Again, however, the eight 
terms we use are many more than the minimum needed.

Our final error estimates depend upon the widths of our 
priors~\cite{Davies:2008sw}. We tested these widths in a couple of 
ways, beyond including simulation data from the coarsest lattices. 
First we compared our widths with the values suggested by the empirical 
Bayes procedure described in~\cite{Lepage:2001ym}. This procedure uses 
the variation in the data itself to determine, for example, an optimal 
value for~$\sigma_c$. The widths we use are two to four times larger 
that what is indicated by the empirical Bayes criterion, suggesting 
that our error estimates are conservative. The dominant fit 
coefficients in the $(am_c)^2$ expansion for $R_6$, for example, range 
between~$-0.05$ and~$-0.20$, which is much smaller than the 
$\sigma_c=1$ we use.

As a second test, we verified that our extrapolation procedure gives 
consistent results when data from either the smallest or the largest 
lattice spacing is discarded. That is, we demonstrated that results 
obtained from the truncated data sets agree within errors with results 
from the full set of simulation data. This shows that our error 
estimates are robust even when working with limited simulation data 
sets. As mentioned above, our final results are not much affected by 
data from the coarsest lattice spacing. Simulation data from the finest 
lattice spacing, on the other hand, has a very significant impact.

\section{Extracting $m_c(\mu)$ and $\alpha_\msb(\mu)$}\label{sec:mc}

\begin{figure}
           
\includegraphics[scale=1.0]{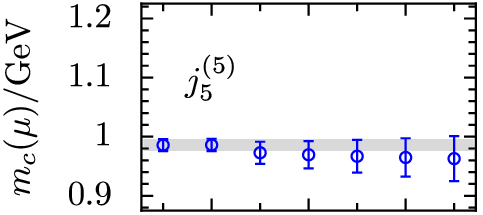}\includegraphics[scale=1.0]{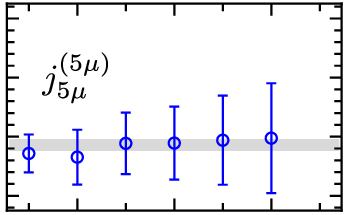}\\
        
\includegraphics[scale=1.0]{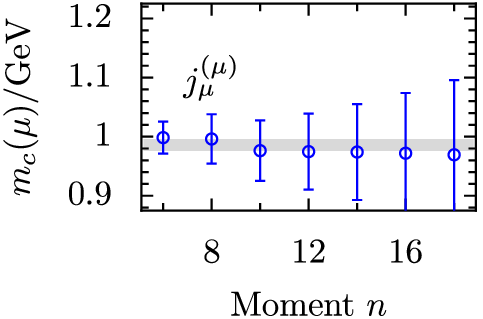}\includegraphics[scale=1.0]{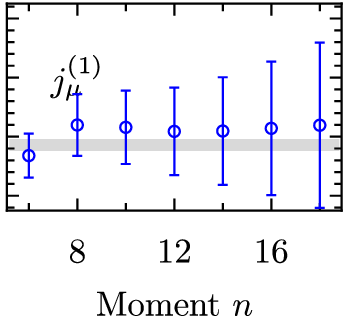}
       \caption{$m_c(\mu)$, for $\mu=3$\,GeV and $n_f=4$ flavors,
	from different moments of correlators built
    from four different lattice operators. The gray band
    is our final result for the mass, $\mcmu$\,GeV, which comes from 
the first two
	moments of the pseudoscalar correlator (upper-left panel).}
    \label{fig:mc}
\end{figure}

To convert the extrapolated reduced moments into $c$~masses and 
coupling constants, we require perturbative expansions for the~$r_n$ in 
\eq{extract-mc}. These are easily computed from the expansions for 
$g_n$~\cite{Chetyrkin:1995ii,Chetyrkin:1996cf,Chetyrkin:1997mb,Boughezal:2006uu,Maier:2007yn,Chetyrkin:2006xg,Boughezal:2006px} using \eq{rn-def}; details can be found in the~Appendix. The perturbative expansions have the form
\begin{equation} \label{rn-exp}
    r_n = 1 + r_{n,1}\alpha_\msb(\mu) + r_{n,2}\alpha_\msb^2(\mu)
    + r_{n,3}\alpha_\msb^3(\mu) + \ldots
\end{equation}
where we set the renormalization scale~$\mu$ 
to~3\,GeV~\cite{footnote:4.5}.
The full third-order coefficients for the $n=4,6,8$ moments were 
computed for this analysis and are presented in the Appendix. The 
third-order coefficients for moments with $n\ge10$ are only partially 
complete: our analysis includes all $\mu$-dependent terms (that is, 
$\log^n(\mu/m_c)$ terms), but the constant parts have not yet been 
computed. Consequently
we take the truncation uncertainty in $r_n$ to be of 
order~\cite{footnote:5}
\begin{equation}\label{rn-err}
    \sigma_{r_n} = \begin{cases}
        r_n^\mathrm{max}\alpha_\msb^4(\mu) & \text{for $n=4,6,8$,} 
\\[2ex]
        r_n^\mathrm{max}\alpha_\msb^3(\mu) & \text{for $n\ge10$,}
    \end{cases}
\end{equation}
where
\begin{equation}
    r_n^\mathrm{max} = \mathrm{max}\left(|r_{n,1}|,|r_{n,2}|,|r_{n,3}| 
\right).
\end{equation}

Another source of uncertainty in all of our moments comes from 
nonperturbative effects. In the previous section, we discuss how we 
remove nonperturbative contributions involving the sea-quark masses. To 
assess the importance of gluonic contributions, we also include the 
leading gluon-condensate contribution in our 
moments~\cite{Novikov:1977dq,Broadhurst:1994qj,Kuhn:2007vp}. We do this 
by multiplying $r_n$ by a factor of the form $(1+d_n\langle \alpha_s 
G^2/\pi\rangle/(2m_c)^4))$ where, here, $m_c=m_c(m_c)$ and $d_n$ is 
computed through leading order in $\alpha_\msb(m_c)$. The value of the 
condensate is not well known; we set $\langle \alpha_s G^2/\pi\rangle = 
0\pm0.012\,\mathrm{GeV}^4$, which covers the range of most current 
estimates~\cite{cond-footnote}.

Note that coefficients in the $r_n$~expansion, \eq{rn-exp}, depend 
upon~$m_c(\mu)$ through scale-dependent 
logarithms,~$\log^n(\mu/m_c(\mu))$. Consequently, the mass appears on 
both sides of~\eq{extract-mc}, and the equation is an implicit equation 
for~$m_c(\mu)$. The $m_c(\mu)$-dependence on the right-hand side, 
however, is suppressed by~$\alpha_\msb(\mu)$, and therefore the 
equation is easily solved numerically.

\begin{figure}
	\begin{center}
		\includegraphics[scale=1.0]{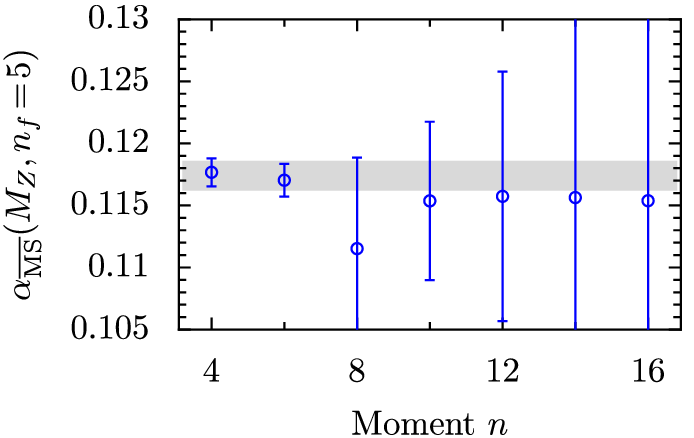}
	\end{center}
	\caption{$\alpha_\msb(M_Z,n_f\!=\!5)$ from $R_4$ and ratios 
$R_n/R_{n+2}$. The gray band is our final result for the coupling, 
\almz, which comes from $R_4$ and $R_6/R_8$.}
	\label{fig:almz}
\end{figure}

Our final results for the $c$-quark's mass, $m_c(\mu)$ at 
$\mu=3\,\mathrm{GeV}$ for $n_f=4$ flavors in the $\msb$~scheme, are 
listed in Table~\ref{tab:rn}, and plotted in the upper-left panel of 
Figure~\ref{fig:mc}. As is clear from the figure, all moments agree on 
the mass although the higher moments may be less trustworthy 
(see~\cite{Kuhn:2007vp}). The first two moments ($n=6,8$) give results 
that are twice as accurate as the others because we have full 
$\order(\alpha_\msb^3)$ perturbation theory in these cases. We average 
the two results, which agree, to obtain our final result for the mass:
\begin{equation} \label{eq:ps-mc}
    m_c(3\,\mathrm{GeV},n_f\!=\!4) = \mcmu\,\mathrm{GeV}
\end{equation}
Evolving down to scale $\mu=m_c(\mu)$ using fourth-order 
evolution~\cite{van 
Ritbergen:1997va,Czakon:2004bu,Chetyrkin:1997dh,Vermaseren:1997fq}, 
this is equivalent to~\cite{mc-footnote}
\begin{equation}
	m_c(m_c,n_f\!=\!4) = \mcmc\,\mathrm{GeV}.
\end{equation}

\begin{table}
\caption{Sources of uncertainty in the determinations
of $m_c(\mu=3\,\mathrm{GeV},n_f\!=\!4)$ and 
$\alpha_\msb(M_Z,n_f\!=\!5)$
from different reduced moments
$R_n$ of the pseudoscalar correlator. The uncertainties
listed are percentages of the final results.}
\label{tab:err}
\begin{center}
\begin{ruledtabular}\begin{tabular}{ldddd}
	& \multicolumn{2}{c}{$m_c(\mu)$} & 
\multicolumn{2}{c}{$\alpha_\msb(M_Z)$} \\
	& \multicolumn{1}{c}{$R_6$}
	& \multicolumn{1}{c}{$R_8$}
	& \multicolumn{1}{c}{$R_4$}
	& \multicolumn{1}{c}{$R_6/R_8$} \\ \hline
                $a^2$ extrapolation &  0.2\% &  0.3\% &  0.4\% &  
0.2\%\\
           perturbation theory &  0.4 &  0.3 &  0.6 &  0.6\\
     $\alpha_\msb$ uncertainty &  0.3 &  0.4 & \tdash & \tdash \\
        $m_c(\mu)$ uncertainty &  \tdash & \tdash &  0.1 &  0.1\\
              gluon condensate &  0.3 &  0.0 &  0.4 &  0.7\\
            statistical errors &  0.1 &  0.0 &  0.2 &  0.1\\
  $m_{0c}$ errors from $r_1/a$ &  0.5 &  0.6 &  0.3 &  0.4\\
    $m_{0c}$ errors from $r_1$ &  0.6 &  0.6 &  0.1 &  0.1\\
     $m_{u/d/s}$ extrapolation &  0.2 &  0.2 &  0.2 &  0.4\\
                 finite volume &  0.1 &  0.1 &  0.0 &  0.3\\
        $\mu\to M_Z$ evolution &  0.0 &  0.0 &  0.1 &  0.1\\
\hline
               Total  &  1.0\% &  1.1\% &  1.0\% &  1.1\%
\end{tabular}\end{ruledtabular}
\end{center}
\end{table}

We used $\alpha_\msb(3\,\mathrm{GeV},n_f\!=\!4)=0.252\,(10)$ in the 
perturbation theory needed to extract~$m_c(\mu)$. We derived this from 
the current Particle Data Group average for the $n_f=5$~coupling at 
$\mu=M_Z$, which is~0.1176\,(20)~\cite{pdg}. The coupling can also be 
extracted directly from $R_4$ and from the ratios $R_n/R_{n+2}$, as 
discussed above. Taking $m_c(\mu)=\mcmu$\,GeV, we obtain the couplings, 
for scale $\mu=3$\,GeV and $n_f=4$, shown in Table~\ref{tab:rn}.  The 
first two determinations listed in the table are far more accurate than 
the others because we know perturbation theory through third order. We 
can average these to obtain a composite value for the coupling of:
\begin{equation}
	\alpha_\msb(3\,\mathrm{GeV},n_f\!=\!4) = \almu .
\end{equation}
To allow comparison with other work we converted our couplings to 
$n_f=5$ by adding a $b$-quark with mass 
$m_b(m_b)=4.20(7)$\,GeV~\cite{pdg}, and evolving them to scale~$M_Z$. 
The results are shown in Figure~\ref{fig:almz}. Averaging the first two 
numbers, which agree with each other, we get:
\begin{equation}
	\alpha_\msb(M_Z,n_f\!=\!5) = \almz .
\end{equation}

The leading sources of uncertainty in $m_c(\mu)$ and $\alpha_\msb(M_Z)$ 
are listed in Table~\ref{tab:err} for those calculations where we have 
full perturbation theory through 
$\order(\alpha_\msb^3)$~\cite{Davies:2008sw}. The dominant uncertainty 
in the masses comes from potential tuning errors in the $c$-quark 
masses used in the simulation. Truncation errors from perturbation 
theory dominate for the coupling, with nonperturbative contributions 
from the gluon condensate also becoming important. In addition to the 
various sources discussed above, there are also uncertainties due to 
the finite spatial volume of our lattices; our lattices were 
approximately 2.5\,fm~across. While our simulations showed no 
measurable volume dependence~\cite{footnote:2.1}, lattice perturbation 
theory shows finite-volume sensitivity for the higher (more infrared) 
moments. This is negligible for lower moments but grows with~$n$. The 
finite-volume sensitivity is mostly an artifact of perturbation theory; 
confinement significantly reduces finite-volume effects. Consequently 
we assign a finite-volume error to our perturbative factors that is 
equal to the entire finite-volume correction in perturbation theory.

\section{$m_c(\mu)$ From Other Correlators}\label{sec:other}
The close agreement on $m_c$ between different moments is important 
evidence that we understand our systematic errors since these enter 
quite differently in different moments. To further check this we 
repeated our analysis for three different correlators, which we formed 
by replacing the pseudoscalar operator $m_{0c}j_5$ with each of the 
following $c$-quark currents on the lattice:
\begin{align}
	j_{\mu}^{(1)} &\equiv \psib_c(x+a\hat\mu)\gamma_\mu\psi_c(x),\\
	j_{\mu}^{(\mu)} &\equiv \psib_c(x)\gamma_\mu\psi_c(x), \\
	j_{5\mu}^{(5\mu)} &\equiv \psib_c(x)\gamma_5\gamma_\mu\psi_c(x).
\end{align}
The first two currents are different lattice discretizations of the 
vector current and were evaluated for space-like $\mu$s; and the first 
of these was evaluated in Coulomb gauge. The third current is a lattice 
discretization of the axial vector current and was evaluated for 
time-like~$\mu$. The superscript on each $j$ labels the ``taste'' 
carried by that operator, using the notation presented in the 
Appendices of~\cite{Follana:2006rc}. Taste is a spurious quantum 
number, analogous to flavor, that is an artifact of staggered-quark 
lattice discretizations like the HISQ formalism. Taste should not 
affect physical results and therefore operators carrying different 
taste here should give identical results in the $a\to0$~limit. By 
studying these different currents, we not only test for conventional 
systematic errors, but also verify that HISQ-specific taste effects are 
negligible~\cite{footnote:6}.

\begin{table}
	\caption{Simulation results for the reduced moments $R_n^{(j)}$, 
extrapolated to~$a=0$,
	from correlators of local axial-vector and vector lattice currents,
	and a point-split lattice
	vector current. Corresponding values for $m_c(\mu)$ (in GeV), for 
$\mu=3$\,GeV and
	$n_f=4$,
	are also given. Only results for parameter sets 1, 4 and~6 from 
Table~\ref{tab:param}
	were used for the first and last currents;  results from these sets 
were combined with
	results from set~8 (the smallest lattice spacing)
	for~$j_\mu^{(\mu)}$. }
	\label{tab:rnj}
	\begin{ruledtabular}
		\begin{tabular}{c|cc|cc|cc} \multicolumn{1}{c}{}
  & \multicolumn{2}{c}{$j_{5\mu}^{(5\mu)}$} &
	\multicolumn{2}{c}{$j_{\mu}^{(\mu)}$} &
	\multicolumn{2}{c}{$j_{\mu}^{(1)}$} \\
\multicolumn{1}{c}{n}
& $R_n^{(j)}$ & \multicolumn{1}{c}{$m_c(\mu)$}
& $R_n^{(j)}$ & \multicolumn{1}{c}{$m_c(\mu)$}
& $R_n^{(j)}$ & $m_c(\mu)$ \\ \hline
 6 & 1.240(27) & 0.97(3) & 1.233(16) & 1.00(3) & 1.261(30) & 0.97(4) \\
 8 & 1.159(25) & 0.97(5) & 1.183(15) & 1.00(4) & 1.163(27) & 1.02(5) \\
10 & 1.126(24) & 0.99(5) & 1.162(15) & 0.98(5) & 1.132(27) & 1.02(6) \\
12 & 1.103(24) & 0.99(6) & 1.139(15) & 0.97(6) & 1.113(26) & 1.01(7) \\
14 & 1.082(23) & 0.99(8) & 1.120(15) & 0.97(8) & 1.094(26) & 1.01(9) \\
16 & 1.064(23) & 1.00(9) & 1.106(14) & 0.97(10) & 1.076(25) & 1.01(11) 
\\
18 &      & & 1.093(14) & 0.97(13) & 1.059(25) & 1.02(14) \\
\end{tabular}

	\end{ruledtabular}
			\end{table}

A complication in our lattice analysis of these vector (or 
axial-vector) correlators is that none of the currents is conserved (or 
partially conserved) on the lattice. Consequently, each lattice current 
is related to its corresponding continuum operator by a renormalization 
constant:
\begin{align}
	j_\mathrm{cont} &= Z^{(j)}\,j \,+\, \order(a^2)  \\
	Z^{(j)} &\equiv Z^{(j)}(\alpha_\msb(\pi/a),am_{0c}) \nonumber
\end{align}
where $j$ is one of the lattice currents $j_\mu^{(1)}$, 
$j_\mu^{(\mu)}$, or $j_{5\mu}^{(5\mu)}$, and $j_\mathrm{cont}$ is the 
continuum current $j_\mu=\psib\gamma_\mu\psi$ for the first two $j$s 
and $j_{5\mu}=\psib\gamma_5\gamma_\mu\psi$ for the last. Consequently 
moments of the correlators of these lattice currents have the form
\begin{equation} \label{Gnv-def}
    G_n^{(j)} = \frac{1}{{Z^{(j)}}^2}\,
		\frac{g_n^{(j_\mathrm{cont})}(\alpha_\msb(\mu),\mu/m_c)}{(am_c(\mu))^{n-2}}.
\end{equation}
where ${Z^{(j)}}^2 G_n^{(j)}$ is the continuum result for $n\ge4$. To 
cancel the renormalization factor we redefine the reduced moments for 
these correlators to be
\begin{align}
	R_n^{(j)} &\equiv \frac{am^{(j)}}{2am_{0c}}\left(
	\frac{G_n^{(j)}}{G_{n-2}^{(j)}} \,
	\frac{G_{n-2}^{(j0)}}{G_n^{(j0)}}
	\right)^{1/2} \\ \label{rnj-def}
	&\equiv 
\frac{r_n^{(j_\mathrm{cont})}(\alpha_\msb,\mu/m_c)}{2m_c(\mu)/m^{(j)}}
\end{align}
where $n\ge6$, and $m^{(j)}$ is the $\psi$~mass for the vector currents 
(which couple to the $\psi$) and the $\eta_c$~mass for the axial-vector 
current. Again we divide each moment $G_n^{(j)}$ by its value 
$G_n^{(j0)}$ in leading-order lattice perturbation theory in order to 
minimize finite-lattice-spacing errors. And again the perturbative 
expansion for $r_n^{(j_\mathrm{cont})}$ can be obtained from continuum 
perturbation theory expansions for the $g_n^{(j_\mathrm{cont})}$ (see 
the Appendix).

Our simulation results for~$R_n^{(j)}$, extrapolated to lattice 
spacing~$a=0$, are given in Table~\ref{tab:rnj} for different 
moments~$n$ and each of the three currents~\cite{footnote:6.5}. 
Perturbative coefficients for the 
vector-current~$r_n^{(j_\mathrm{cont})}$s are discussed in the 
Appendix; the coefficients for the temporal axial-vector current can be 
derived from the pseudoscalar coefficients (also in the Appendix) using 
Ward identities~\cite{footnote:7}.

By combining perturbative with nonperturbative results, we obtain the 
values for $m_c(\mu)$, with $\mu=3$\,GeV and $n_f=4$, that are listed 
in Table~\ref{tab:rnj} and plotted in the top-right and bottom panels 
of Figure~\ref{fig:mc}. Results from all moments agree with each other 
and with the pseudoscalar result (the gray band in the plots), although 
here the errors are about twice as large for the smaller moments.

\begin{figure}
	\begin{center}
		\includegraphics{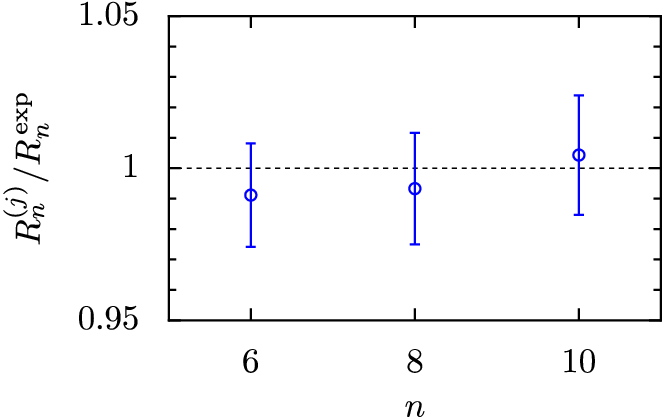}
	\end{center}
	\caption{Ratio of the extrapolated simulation results from 
Table~\ref{tab:rnj}
	for $R_n^{(j)}$, with
	$j=j^{(\mu)}_\mu$, to results derived from experiment 
in~\cite{Kuhn:2007vp} for
	different moments~$n$.}
	\label{fig:rn-ratio}
\end{figure}

Values for the lowest four moments of the vector correlator are derived 
from experimental data for $e^+e^-$~annihilation in~\cite{Kuhn:2007vp}. 
Converting these into reduced moments, we compare them with our 
extrapolated $R^{(j)}_n$s for $j=j^{(\mu)}_\mu$ (from 
Table~\ref{tab:rnj}) in Figure~\ref{fig:rn-ratio}. We find that 
experiment and our simulation results agree to within combined errors 
of better than~$2\%$. This comparison is more accurate than comparing 
masses, because we are comparing the (reduced) moments directly, 
without recourse to further perturbation theory.

\begin{figure}
           \includegraphics[scale=1.0]{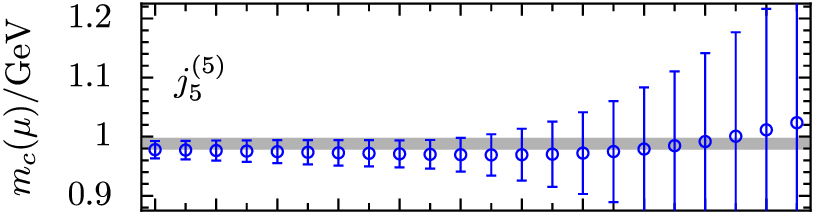}\\
        \includegraphics[scale=1.0]{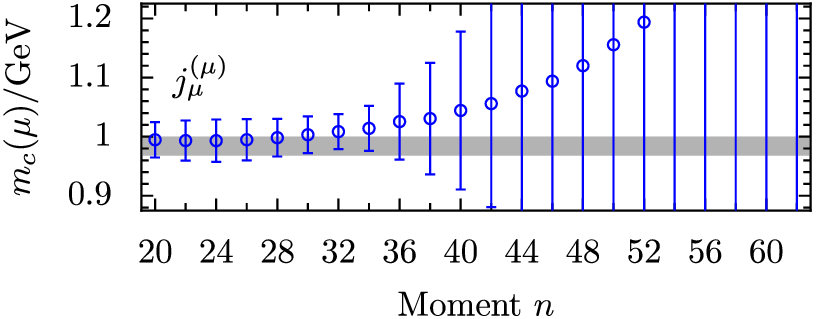}
       \caption{$m_c(\mu=3\,\mathrm{GeV})$ from large-$n$ moments of
     the pseudoscalar and the (local) vector correlators. The gray band
    is our final result for the mass (from $n=6,8$), \mcmu\,GeV.
	The perturbative part of the
	analysis was evaluated at $\mu=m_c(\mu)$ using formulas 
from~\cite{Maier:2007yn}, and
	the results evolved to $\mu=3$\,GeV using fourth-order evolution. 
Uncertainties due
	to the gluon condensate are not included (see text).}
    \label{fig:mc-largen}
\end{figure}

\section{Conclusions}\label{concl}
Our results are by far the most accurate determination of the $c$-quark 
mass from lattice QCD~\cite{old-lattice-mc}. Such precision is possible 
because the matching between lattice parameters and continuum 
parameters here relies upon continuum perturbation theory, which is 
much simpler than lattice QCD perturbation theory. Consequently 
perturbation theory can be pushed to much higher orders. The precision 
of this continuum calculation is matched by that of our nonperturbative 
lattice analysis because of the quality of the MILC configurations and 
of our highly corrected HISQ action for $c$-quarks.

The agreement between masses from different moments, and from different 
correlators\,---\,27 determinations in all\,---\,is an important check 
on systematic errors of all sorts since these enter in very different 
ways in each calculation. Note that the different reduced moments in 
our analysis vary in value by as much as 43\% (from 1.06 to 1.52), and 
yet they all agree on the value of~$m_c$ to within a few percent once 
we account for differences in the perturbative parts.

One surprising feature of our results is that even the higher moments 
give correct values for the quark mass, albeit with larger errors. 
Nonperturbative effects grow with~$n$ but our results show no 
systematic deviation until very large~$n$, as is evident in 
Fig.~\ref{fig:mc-largen} which shows results for~$20\le n\le62$. We 
have not included potential errors due to the gluon condensate in this 
figure. The error bars would have been much larger had we done so. For 
example, they would have been about 5~times larger at $n=40$ in the 
pseudoscalar plot (16\%~rather than~3\%). This might suggest that the 
condensate is smaller than we allowed for\,---\,say $\langle \alpha_s 
G^2/\pi\rangle\le0.003\,\mathrm{GeV}^4$\,---\,but we have not analyzed 
this carefully enough to make a strong statement. The error bars shown 
in the plots start to grow rapidly just where it becomes clear that 
perturbation theory is failing (because of large coefficients).

Our lattice result for the mass, 
$m_c(3\,\mathrm{GeV},n_f\!=\!4)=\mcmu$\,GeV, agrees well with the 
continuum determination from $e^+e^-$~annihilation data, which gives 
0.986\,(13)\,GeV~\cite{Kuhn:2007vp}. This provides further strong 
evidence that the different systematic errors in each calculation are 
understood. Similarly our new value for the coupling, 
$\alpha_\msb(M_Z,n_f\!=\!5)=\almz$, agrees very well with non-lattice 
determinations~\cite{pdg,Bethke:2006ac} and our other determinations 
from lattice QCD~\cite{Mason:2005zx,Davies:2008sw}. It is also more 
accurate than most determinations.

The close agreement of our results with non-lattice determinations of 
the mass and coupling, and the taste-independence of our masses, is 
also further evidence that the simulation methods we use are valid. 
While early concerns about the light-quark discretization used here 
have been largely addressed~\cite{Sharpe:2006re,Bernard:2007eh}, it 
remains important to test the simulation technology of lattice QCD at 
increasing levels of precision, given the critical importance of 
lattice results for phenomenology.

Our results are particularly relevant to the recent, very accurate 
analysis of $\pi$, $K$, $D$ and $D_s$ meson decay constants using the 
same HISQ formalism for valence quarks and many of the same MILC 
configuration sets that we use here~\cite{Follana:2007uv}. The 
correlators in our pseudoscalar analysis are identical to those used to 
extract the decay constants in the earlier study, except that here the 
light valence quarks have been replaced by $c$~quarks, which should 
make finite-$a$ errors worse. The agreement of our pseudoscalar 
analysis of $R_6$ with the continuum analysis for $m_c(\mu)$ indicates 
that our extrapolated lattice  $c\overline{c}$~correlators are reliable 
to within~2\% or better. The accuracy of our $c\overline{c}$ results, 
together with the demonstrated accuracy of the $\pi$ and $K$ results 
in~\cite{Follana:2007uv} from light-quark correlators, strongly suggest 
that the corresponding $D$ and $D_s$ predictions, from correlators with 
a light quark and a $c$~quark, are reliable. This conclusion is made 
more important by the $3.6\sigma$ discrepancy between the 
$D_s$~prediction and recent experimental 
results~\cite{cleo:2007zm,Artuso:2007zg,Aubert:2006sd}.

The lattice analysis will be improved as data becomes available for 
smaller lattice spacings. Also a very accurate $c$-quark mass will 
allow us to make similarly accurate determinations of the $s$-quark 
mass~\cite{c-s-ratio}. This is because the ratio $m_s/m_c$ can be 
determined very accurately in lattice simulations where the $s$~and 
$c$~quarks are analyzed using the same formalism, as here. Finally 
four-loop perturbation theory for additional moments would improve both 
lattice and continuum determinations.

\section*{Acknowledgements}
We would like to thank A. Maier, P. Maierh\"ofer and P. Marquard
for providing $\bar{C}^{(30)}_3$ for the pseudo-scalar correlator prior 
to publication. We are grateful to the MILC Collaboration for the use 
of their gluon configurations.
We thank Rainer Sommer for useful discussions.
The lattice QCD computing was done on UKQCD's QCDOCX cluster, USQCD's 
Fermilab cluster, and at the Ohio Supercomputer Center. The work was 
supported by grants from: the Deutsche Forschungsgemeinschaft, 
SFB-Transregio 9; the Department of Energy (DE-FG02-91-ER40690, 
DE-AC02-98CH10886(BNL)); the Leverhulme Trust; the Natural Sciences and 
Engineering Research Council; the National Science Foundation; and the 
Science and Technology Facilities Council.

\newpage

\begin{widetext}
\begin{center}
\begin{table}[ht]
\begin{center}
{\begin{tabular}{r|rrrrrrrrrr}
\hline
\hline
 $k$ &
 $\bar{C}_k^{(0)}$  & $\bar{C}_k^{(10)}$ & $\bar{C}_k^{(11)}$ &
 $\bar{C}_k^{(20)}$ & $\bar{C}_k^{(21)}$ & $\bar{C}_k^{(22)}$ &
 $\bar{C}_k^{(30)}$ & $\bar{C}_k^{(31)}$ & $\bar{C}_k^{(32)}$ &
 $\bar{C}_k^{(33)}$
 \\
 \hline
1&$   1.3333$&$   3.1111$&$   0.0000$&$   0.1154$&$  -6.4815$&$   
0.0000$&$  -1.2224$&$   2.5008$&$  13.5031$&$   0.0000$\\
2&$   0.5333$&$   2.0642$&$   1.0667$&$   7.2362$&$   1.5909$&$  
-0.0444$&$   7.0659$&$  -7.5852$&$   0.5505$&$   0.0321$\\
3&$   0.3048$&$   1.2117$&$   1.2190$&$   5.9992$&$   4.3373$&$   
1.1683$&$$ 14.5789$$&$   7.3626$&$   4.2523$&$  -0.0649$\\
4&$   0.2032$&$   0.7128$&$   1.2190$&$   4.2670$&$   4.8064$&$   
2.3873$&$$---$$&$  14.7645$&$  11.0345$&$   1.4589$\\
5&$   0.1478$&$   0.4013$&$   1.1821$&$   2.9149$&$   4.3282$&$   
3.4971$&$$---$$&$  16.0798$&$  16.6772$&$   4.4685$\\
6&$   0.1137$&$   0.1944$&$   1.1366$&$   1.9656$&$   3.4173$&$   
4.4992$&$$---$$&$  14.1098$&$  19.9049$&$   8.7485$\\
7&$   0.0909$&$   0.0500$&$   1.0912$&$   1.3353$&$   2.2995$&$   
5.4104$&$$---$$&$  10.7755$&$  20.3500$&$  14.1272$\\
8&$   0.0749$&$  -0.0545$&$   1.0484$&$   0.9453$&$   1.0837$&$   
6.2466$&$$---$$&$   7.2863$&$  17.9597$&$  20.4750$\\
\hline
1&$   1.0667$&$   2.5547$&$   2.1333$&$   2.4967$&$   3.3130$&$  
-0.0889$&$  -5.6404$&$   4.0669$&$   0.9590$&$   0.0642$\\
2&$   0.4571$&$   1.1096$&$   1.8286$&$   2.7770$&$   5.1489$&$   
1.7524$&$ -3.4937$&$   6.7216$&$   6.4916$&$  -0.0974$\\
3&$   0.2709$&$   0.5194$&$   1.6254$&$   1.6388$&$   4.7207$&$   
3.1831$&$$---$$&$   7.5736$&$  13.1654$&$   1.9452$\\
4&$   0.1847$&$   0.2031$&$   1.4776$&$   0.7956$&$   3.6440$&$   
4.3713$&$$---$$&$   4.9487$&$  17.4612$&$   5.5856$\\
5&$   0.1364$&$   0.0106$&$   1.3640$&$   0.2781$&$   2.3385$&$   
5.3990$&$$---$$&$   0.9026$&$  18.7458$&$  10.4981$\\
6&$   0.1061$&$  -0.1158$&$   1.2730$&$   0.0070$&$   0.9553$&$   
6.3121$&$$---$$&$  -3.1990$&$  16.9759$&$  16.4817$\\
7&$   0.0856$&$  -0.2033$&$   1.1982$&$  -0.0860$&$  -0.4423$&$   
7.1390$&$$---$$&$  -6.5399$&$  12.2613$&$  23.4000$\\
8&$   0.0709$&$  -0.2660$&$   1.1351$&$  -0.0496$&$  -1.8261$&$   
7.8984$&$$---$$&$  -8.6310$&$   4.7480$&$  31.1546$\\
\hline
\hline
\end{tabular}
}
\end{center}
\caption{\label{tab:1}
Moments of the pseudoscalar
(upper 8 lines) and the vector (lower 8 lines) correlators, where 
currently unknown coefficients are denoted by a dash and set to zero in 
our analysis.
The numbers correspond  to QCD with one  massive $c$-quark and three 
massless $(u,d,s)$ quarks.
}
\end{table}
\end{center}
\end{widetext}

\section*{Appendix A: Continuum perturbation 
theory\label{app:PT}}\label{cont-pert-th}
The correlators of two pseudoscalar ($\imath\, 
\bar{\psi_c}\g_5\psi_c$),
vector $\bar{\psi_c}\g_{\mu}\psi_c$, and
axial-vector $\bar{\psi_c}\g_{\mu}\g_5\psi_c$ currents are defined by
\begin{equation}
\label{eq:pseudoscalar}
q^2\*\Pi^{p}(q^2)=i\int\!dx e^{iqx}\langle0|Tj_5(x)j_5(0)|0\rangle,
\end{equation}
\begin{eqnarray}
\label{eq:axialvector}
(q_{\mu}q_{\nu}-q^2g_{\mu\nu})\*\Pi^{\delta}(q^2)
 +q_{\mu}q_{\nu}\Pi^{\delta}_{L}(q^2)=\Pi^{\delta}_{\mu\nu}(q)
\nonumber\\
=i\int\!dx\, 
e^{iqx}\langle0|Tj_{\mu}^{\delta}(x)j_{\nu}^{\delta}(0)|0\rangle
{}.
\end{eqnarray}
where $\delta=v$ and~$a$ for the vector and axial-vector cases, 
respectively.
The polarization functions can be expanded in the variable
$z=q^2/(2\*m_c(\mu))^2$:
\begin{equation}
\label{eq:expand}
\bar{\Pi}^{\delta}(q^2)=\frac{3}{16\pi^2}\sum_{k=-1}^{\infty}\bar{C}^{\delta}_k\*z^k
{}.
\end{equation}
The polarization function and the $c$-quark mass $m_c(\mu)$ are 
renormalized
in the $\overline{\mbox{MS}}$-scheme.

The longitudinal part of the axial-vector current (which is of interest
in the present context) and the pseudoscalar correlator are related by
the axial Ward-identity~\cite{Broadhurst:1981jk}:
\begin{equation}
\label{eq:AWI}
q^\mu\*q^\nu\*\Pi^a_{\mu\nu}(q)=(2\*m)^2\*q^2\*\Pi^p(q^2)+\mbox{contact 
term.}
\end{equation}
Comparing different orders in $z$,
\begin{equation}
\label{eq:AWIExp}
\bar{C}^p_{k+1}=\bar{C}^{a}_{L,k},\quad\mbox{for}\quad k\ge-1,
\end{equation}
allows us to extract moments of the pseudoscalar correlator from those
of the longitudinal part of the axial-vector correlator, and \emph{vice 
versa}.
The contact term  contributes only to  $k=-2$.
The coefficients of the perturbative expansion depend logarithmically 
on
$m_c$ and can be written in the form
\begin{eqnarray}
  \bar{C}_k &=&
  \bar{C}_k^{(0)}
  + \frac{\alpha_s(\mu)}{\pi}
  \left( \bar{C}_k^{(10)} + \bar{C}_k^{(11)}\lmc \right) \nonumber \\
  &+& \left(\frac{\alpha_s(\mu)}{\pi}\right)^2
  \left( \bar{C}_k^{(20)} + \bar{C}_k^{(21)}\lmc
  + \bar{C}_k^{(22)}\lmc^2 \right)
  \nonumber\\
  &+& \left(\frac{\alpha_s(\mu)}{\pi}\right)^3
  \left( \bar{C}_k^{(30)} + \bar{C}_k^{(31)}\lmc \right. \nonumber \\
  &&\quad\quad\quad\quad\quad\left. + \bar{C}_k^{(32)}\lmc^2 + 
\bar{C}_k^{(33)}\lmc^3 \right)
  + \ldots
  \,.
  \label{eq::cn}
\end{eqnarray}
where $l_{m_c}\equiv \log(m_c^2(\mu)/ \mu^2)$. The coefficients
$\bar C_k^{(ij)}$ up through $k=8$ are listed in Table~\ref{tab:1} for 
both pseudoscalar
and vector correlators.

The four-loop coefficients $\bar C_1^{(30)}$ and $\bar C_2^{(30)}$
for the pseudoscalar correlator are new~\cite{Sturm:2008aa}; and
$\bar C_1^{(30)}$ for the vector correlator comes 
from~\cite{Chetyrkin:2006xg,Boughezal:2006px}. The four-loop 
coefficients $\bar C_3^{(30)}$ for the pseudoscalar case is also 
new~\cite{Maier:private}, as is $\bar C_2^{(30)}$ for the vector 
case~\cite{Maier:2008he}.
The order $\alpha_s^2$ terms up through $k=8$ are given in 
\cite{Chetyrkin:1995ii,Chetyrkin:1996cf,Chetyrkin:1997mb},
while results for higher $k$-values are given in
\cite{Boughezal:2006uu,Maier:2007yn}. See also~\cite{Czakon:2007qi} for 
$n_f$-dependent four-loop terms, and~\cite{Maier:2007yn} for the 
pseudoscalar case.  Throughout this paper we take the number of light 
(massless) active quark flavors to be $n_l=3$, and
the number of heavy (massive) quarks is set to $n_h=1$. For numerical 
work, we use~$\mu=3\,$GeV.

The expansion coefficients $\bar{C}_k$ are related to the
coefficients $g_n$ of \eq{gn-def} by:
\begin{equation}
\frac{g_{2k+2}}{g_{2k+2}^{(0)}} =
\frac{\bar{C}_{k}}{\bar{C}_{k}^{(0)}},
\label{eq:CngnRelation}
\end{equation}
and the coefficients~$r_{n,i}$ in \eq{rn-exp}
are  obtained  through the series expansion  of \eq{rn-def}.
For the vector correlator  the coefficients  $r^{(j_{\mu})}_{n,i}$
are defined through  the series expansions of
\begin{equation}
r^{(j_{\mu})}_{2k+2} =
\left(
\frac{\bar{C}_k^{v} }{\bar{C}_k^{v,(0)}}
\,
\frac{\bar{C}_{k-1}^{v,(0)}}{\bar{C}_{k-1}^{v} }
\right)^{1/2}
{}.
\end{equation}

\bibliographystyle{plain}

\begin{thebibliography}{16}
\bibitem{lepage:pcac} The HISQ action, like all staggered-quark 
discretizations (but unlike many other discretizations), has an exact 
chiral symmetry in the limit of massless of quarks. The cutoff 
independence follows from the corresponding PCAC relationship.
\bibitem{footnote:0} The value for $t/a$ at the middle of the lattice 
was
set to~0 in order to preserve the symmetry between~$t$ and~$-t$. This 
choice
causes odd moments to vanish, for example. The value chosen for this 
point doesn't
matter in our analysis because contributions to our moments from this 
point are
negligibly small (because of the exponential falloff in the correlators 
at large~$t$).
\bibitem{lepage:tip} The leading-order propagator $G^{(0)}(t)$ is easy 
to compute with the lattice code used to generate $G(t)$: simply 
recompute the correlator with the same code and lattice parameters, but 
replacing the gluon link matrices by the unit matrix everywhere. 
Moments of this gluon-free propagator are the $G_n^{(0)}$.
\bibitem{footnote:1} We use the $\eta_c$ mass because it is the easiest 
charmonium mass to measure accurately in a simulation, and because the 
experimental value is known to better than~0.1\%.
\bibitem{footnote:2.2} We computed the contribution from $c$-quark 
loops by comparing results from
\cite{Maier:2007yn} for $n_h=1$ (with $c$ quarks) with those for 
$n_h=0$ (without $c$ quarks).
The reduced moments~$R_n$ are corrected, to include $c$~quarks, by 
multiplying them
by $r_n(n_h=1)/r_n(n_h=0)$.
\bibitem{footnote:2.1} Note that configuration sets 3, 4 and~5 all have 
roughly the same lattice spacing, but the lattice for set~3 is 20\% 
longer in each spatial direction than the others. Results from set~3 
show no significant differences from those coming from the other two 
sets.
\bibitem{footnote:1.1} One might worry about simulation errors in the 
calculation of $m_{\eta_c}$ itself and their effect on $m_{0c}$. While 
statistical errors in $m_{\eta_c}$ are negligible, there are certainly 
finite-$a$ errors and some (weak) dependence on the sea-quark masses. 
These will induce additional finite-$a$ and sea-quark mass dependence 
in other quantities, like our reduced moments, that depend 
upon~$m_{0c}$. These additional terms are small, because our HISQ 
$c$-quark action is highly improved, and they are handled properly by 
the $a^2$ and chiral analysis of the reduced moments and other 
quantities of interest.
\bibitem{footnote:1.2} In most lattice simulations one wants to tune 
the sea-quark masses to their physical values. Here, however, we 
extrapolate to zero mass in order to remove nonperturbative 
contributions proportional to the sea-quark masses. This extrapolation 
turns out to be almost negligible.
\bibitem{footnote:4.5} An alternative choice, $\mu=m_c(m_c)$, gives 
smaller perturbative
coefficients for the moments of interest, but $\alpha_\msb(\mu)$ is 
substantially larger.
Consequently $\mu=3\,$GeV gives smaller errors than $\mu=m_c(m_c)$ 
despite
having somewhat larger
expansion coefficients. Differences in the final results are small, 
however.
\bibitem{footnote:5} This gives a perturbative error that is
approximately the same size as the error estimates obtained
in~\cite{Kuhn:2007vp} by varying the renormalization scale 
between~$\mu=2$~and~4\,GeV.
The value for $R_6$, for example, varies from $+0.5$\% to $-0.4$\% of 
the central
value as $\mu$ is varied over this range; the error estimate 
from~\eq{rn-err} is~$\pm 0.4$\%.
\bibitem{mc-footnote} Note that the error in $m_c(m_c)=\mcmc$\,GeV is 
smaller than
that in $m_c(\mu)=\mcmu$\,GeV. This is because the scale in $m_c(m_c)$ 
is
not fixed but rather varies as $m_c(m_c)$ itself is varied across the
interval~$\mcmcpm$\,GeV.
\bibitem{footnote:6} A related problem is that correlators built from 
these currents
are contaminated by contributions from correlators built from 
opposite-parity
operators (see Appendix~G of~\cite{Follana:2006rc}). These lattice 
artifacts
oscillate in sign, as $(-1)^{t/a}$, and so are
suppressed by $(am_c/\pi)^{n-2}$ in the $n^\mathrm{th}$~moment since 
the
sum over~$t$ is dominated by~$E\approx0$. Such
contamination is absent in the case of the pseudoscalar correlators.
\bibitem{footnote:6.5} We fit these reduced moments the same way we fit 
the pseudoscalar moments except that we need a prior $\sigma_c=2$ for 
the $c_{n,i}$ that is twice as wide.
\end{thebibliography}

\end{document}